\documentclass[aps, prd,twocolumn,superscriptaddress, nofootinbib, 
showpacs, preprintnumbers]{revtex4}
\usepackage[dvipdfmx]{graphicx,color}
\usepackage{amsmath,amssymb}
\usepackage{bm}
\usepackage{booktabs}
\usepackage{comment}
\usepackage{ulem}

\def\fsl#1{\setbox0=\hbox{$#1$}                 % set a box for #1
 \dimen0=\wd0                                 % and get its size
 \setbox1=\hbox{/} \dimen1=\wd1               % get size of /
 \ifdim\dimen0>\dimen1                        % #1 is bigger
    \rlap{\hbox to \dimen0{\hfil/\hfil}}      % so center / in box
    #1                                        % and print #1
 \else                                        % / is bigger
    \rlap{\hbox to \dimen1{\hfil$#1$\hfil}}   % so center #1
    /                                         % and print /
 \fi}                                         %

\begin{document}
\preprint{MISC-2017-03, IPMU17-0070}
\title{Higgs mediated CLFV processes $\mu N(eN)\rightarrow\tau X$ 
via gluon operators}
\date{\today}
\pacs{11.30.Hv, 13.60.Hb, 14.60.Fg, 14.80.Bn}

\author{Michihisa Takeuchi}
\email[E-mail: ]{michihisa.takeuchi@ipmu.jp}
\affiliation{
Kavli IPMU (WPI), The University of Tokyo, Kashiwa, Chiba 277-8583, Japan
}
\author{Yuichi Uesaka}
\email[E-mail: ]{uesaka@kern.phys.sci.osaka-u.ac.jp}
\affiliation{
Department of Physics, Osaka University, Toyonaka, Osaka 560-0043, Japan
}
\author{Masato Yamanaka}
\email[E-mail: ]{masato.yamanaka@cc.kyoto-su.ac.jp}
\affiliation{
Maskawa Institute, Kyoto Sangyo University, Kyoto 603-8555, Japan
}

\begin{abstract} 
We revisit charged lepton flavor violating (CLFV) scattering processes
$\ell_{i} N \to \tau X \, (\ell_{i} \ni e, \mu)$ mediated by Higgs. 
%%%
We point out that a new subprocess $\ell_{i} g \to \tau g$ via 
the effective interactions of Higgs and gluon
%generated by heavy quark loops.  
%for the scattering $l_{i} N \to \tau X$. 
gives the dominant contribution to $\ell_{i} N \to \tau X$ for an incident beam energy of 
$E_{\ell} \lesssim 1\,\text{TeV}$ in fixed target experiments. 
%%%
Furthermore, in the light of quark number conservation, 
%and of the generation of sea quarks through gluon splitting, 
we consider quark pair-production processes 
$\ell_{i} g \to \tau q \bar{q}$ ($q$ denotes quarks)
instead of $\ell_{i} q \to \tau q$. 
This corrects the threshold energy of each subprocess contributing to $\sigma(\ell_{i} N \to \tau X)$.
%%%
Reevaluation of $\sigma(\ell_{i} N \to \tau X)$ including all 
of relevant subprocesses shows that the search for 
$\ell_{i} N \to \tau X$ could serve a complementary opportunity with other relevant processes
to shed light on the Higgs CLFV. 
\end{abstract}

\maketitle

%%%%%%%%%%%%%%%%%%%%%%
%%%%%%%%%%%%%%%%%%%%%%
%%%%%%%%%%%%%%%%%%%%%%
\section{Introduction}
\label{Sec:int}
%%%%%%%%%%%%%%%%%%%%%%
%%%%%%%%%%%%%%%%%%%%%%
%%%%%%%%%%%%%%%%%%%%%%

One of the most puzzling issues in particle physics is the flavor 
sector; why there are three generations, what the origin of 
mass hierarchy is, and so on. 
%%%
Many types of solutions for this puzzle have been proposed in 
UV completions of the standard model (SM), 
%its examples include
%e.g., two Higgs doublet models, composite Higgs models, 
%sterile neutrino models, extra dimension models
and in general predict a misaligned Yukawa couplings in the mass basis which 
give rise to flavor violating interactions of Higgs and fermions. 
The Higgs field therefore could be a promising probe to the puzzle.

Charged lepton flavor violation (CLFV) is 
an important process to search for the Higgs induced flavor 
violation due to its high sensitivity and a 
variety of observables~\cite{Goudelis:2011un, Harnik:2012pb}. 
We focus on the CLFV involving taus as they are relatively 
less constrained and a sizable effect could be naturally expected. 
Besides that an interesting deviation in $h\to \tau\mu$~\cite{Khachatryan:2015kon} 
also motivates us to focus on the tau CLFV. 
%%%
Once one of the tau CLFV processes is discovered, e.g., 
$\tau \to \ell_{i} \gamma$, 
$\tau \to \ell_{i} \ell_{j} \ell_{k}$, 
$h \to \tau \ell_{i}$, etc., where $i, j, k$ 
denote charged lepton flavor indices, other 
CLFV processes and their correlation should be 
studied for shedding light on the UV structure responsible for 
the CLFV interactions.
%%%
Among the complementary reactions the tau production 
$\ell_{i} N \to \tau X$ 
is relatively less attention paid to. 
Here $N$ is a nucleus and this process is expected to happen 
in fixed target experiments at a sizable rate $\propto \rho$, 
where $\rho$ being target density~\cite{Gninenko:2001id, 
Sher:2003vi}. 
%%%
The beam intensity is planned to reach at $10^{22}$ electrons 
per year in ILC~\cite{Baer:2013cma}, 
and at $\gtrsim 10^{20}$ muons per year in neutrino 
factories and in the muon collider experiment~\cite{Delahaye:2013jla}. 
%%%
Furthermore, LHeC and its optional plans facilitate the electron-proton 
collision at the center of mass 
energy of $\sqrt{s} \gtrsim 1\,\text{TeV}$ 
with the 
luminosity $\mathcal{L} \gtrsim 10^{33}\, 
\text{cm}^{-2} \text{s}^{-1}$~\cite{Acar:2016rde}. 
%%%
In the light of these upcoming experiments, $\ell_{i} N \to \tau X$ 
would be a promising probe for the Higgs induced CLFV.

In this letter, we point out that the gluon initiated partonic 
subprocess, having not considered in the literatures,
%%%
\begin{equation}
\begin{split}
	\ell_{i} g \to \tau g,
\label{Eq:lgtotg}
\end{split}
\end{equation}
provides a dominant contribution to $\ell_{i} N \to \tau X$ in 
fixed target experiments for incident beam energy of $E_{l} 
\lesssim 1\,\text{TeV}$. 
%%%
Furthermore, we stress the importance of 
quark number 
conservation. Since sea quarks are generated through gluon 
splitting, we have to include the final state with a quark pair, 
%%%
 \begin{equation}
\begin{split}
	\ell_{i} g \to \tau q \bar{q}, 
\label{Eq:lgtotqq}
\end{split}
\end{equation}
%%%
instead of sea quark single-production $\ell_{i} q \to \tau q$. 
Here $q=s, c, b, t$. 
%%%
The related subprocess $\ell_{i} q \to \tau q$ has been studied in 
Ref.~\cite{Kanemura:2004jt} in the context of a supersymmetric extension of 
SM, where they consider only the effective 4-Fermi 
operators involving the sea quarks and evaluate the cross sections 
assuming massless patrons. 
%%%
The difference of the available phase space between $\ell_{i} g \to \tau q \bar{q}$ 
and $\ell_{i} q \to \tau q$ is not negligible at the fixed target 
experiments with the relatively low collision energy $\sqrt{s} \simeq 
\sqrt{2ME_{l}} \simeq  13.7 \,\text{GeV} \sqrt{E_{l}/100\,\text{GeV}}$, 
where $M$ denotes the nucleon mass.
%%%
For example, production of $\tau b\bar{b}$ is kinematically 
allowed only when the beam energy exceeds $E_{\ell} \gtrsim 
55\,\text{GeV}$,
while it would be $E_{\ell} 
\gtrsim 19\,\text{GeV}$ when simply the $\tau b$ threshold is considered. 
%%%
In this letter, we reformulate the Higgs mediated tau production 
$\ell_{i} N \to \tau X$ by taking the new effects:
(i) the effective interaction of Higgs and gluons, and
(i\hspace{-0.5pt}i) the 
quark number conservation, 
into account. 
%%%
Due to the altered formulations, the cross section of $\ell_{i} N \to 
\tau X$ drastically changes from the previous estimation for the beam 
energy $E_{\ell} \lesssim 1\,\text{TeV}$. It would largely affect the 
search for this process in the next generation experiments.

%%%%%%%%%%%%%%%%%%%%%%
%%%%%%%%%%%%%%%%%%%%%%
%%%%%%%%%%%%%%%%%%%%%%
\section{CLFV scattering $\ell_{i} N \to \tau X$ with gluon operators} 
\label{Sec:liNtoljX}
%%%%%%%%%%%%%%%%%%%%%%
%%%%%%%%%%%%%%%%%%%%%%
%%%%%%%%%%%%%%%%%%%%%%

The relevant Lagrangian terms for the Higgs mediated CLFV scattering 
$\ell_{i} N \to \ell_{j} X$ are shown in Eq.~(\ref{Eq:Lag}). 
%%%
\begin{equation}
\begin{split}
  &\mathcal{L}
  =
  \mathcal{L}_\text{SM} + \mathcal{L}_\text{CLFV},
  \\
  &\mathcal{L}_\text{SM}
  =
  %g_{s} \bar{q} t^{a} g_{\mu}^{a} \gamma^{\mu} q
  %+
  -\sum_{q} y_{q} h \bar{q} q
  + g_{hgg} h G_{\mu \nu}^{a} G^{a \mu \nu},
  \\
  &\mathcal{L}_\text{CLFV}
  =
  -\rho_{ij} \bar{\ell}_{j} P_{L} \ell_{i} h
  - \rho_{ji} \bar{\ell}_{j} P_{R} \ell_{i} h
\label{Eq:Lag}
\end{split}
\end{equation}
%%%
where $G_{\mu \nu}^{a}$ is gluon field strength, and $g_{hgg}$
is an effective coupling. In the literatures, the scattering is described only
by sea quark contributions $\ell_{i} q \to \tau q$ 
through the quark Yukawa interactions. 
%%%
We investigate the effects of the 
effective operator of Higgs and gluons given by the 
second term and the effects of the quark number conservations. 
CLFV interactions 
are parametrized by the couplings $\rho_{ij}$, where $i$ and $j$ 
are flavor indices of charged leptons, and $i \neq j$. 
%%%
Current bounds on the couplings come from the measurements of 
CLFV Higgs decays at the LHC~\cite{CMS:2016qvi, Aad:2016blu, 
Khachatryan:2016rke},  and are $\sqrt{|\rho_{e \tau}|^2+ 
|\rho_{\tau e}|^2}< 2.4 \times 10^{-3}$ and 
$\sqrt{|\rho_{\mu \tau}|^2+ |\rho_{\tau \mu}|^2}< 
3.16 \times 10^{-3}$, where an interesting excess was 
reported~\cite{Khachatryan:2015kon}.

\begin{figure}[t]
\begin{center}
\includegraphics[width=80mm]{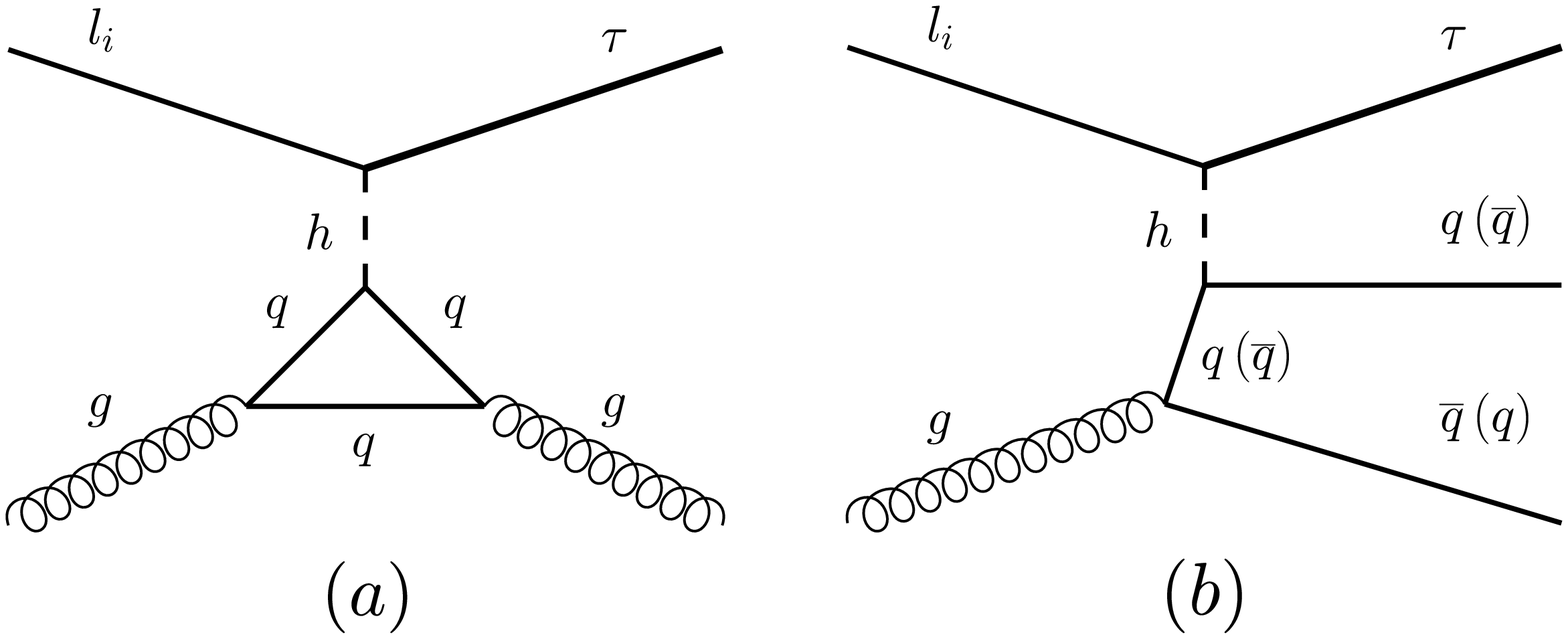}
\end{center}
\vspace{-0.5cm}
\caption{Subprocesses of $\ell_{i} N \to \tau X$.
$(a)$ $\ell_{i} g \to \tau g$ via gluon operator.
$(b)$ $\ell_{i} g \to \tau q \bar{q}$.}
\label{Fig:diagram}
\end{figure}

%%%%%%%%%%%%%%%%%%%%%%
%%%%%%%%%%%%%%%%%%%%%%
%%%%%%%%%%%%%%%%%%%%%%
\subsection{effective gluon Higgs operator}
\label{Sec:ghgg}
%%%%%%%%%%%%%%%%%%%%%%
%%%%%%%%%%%%%%%%%%%%%%
%%%%%%%%%%%%%%%%%%%%%%

The effective coupling $g_{hgg}$ is generated by the triangle 
diagrams where quarks are running (see Fig.~\ref{Fig:diagram} 
(a)), and is derived as a function of the momentum transfer of 
the Higgs $q_{h}$ ($q^2_h =-Q^2 < 0$) as 
follows~\cite{Georgi:1977gs, Spira:1995rr},
%%%
\begin{equation}
\begin{split}
  g_{hgg}
  &=
  \sum_{q=c,b,t} 
\frac{\alpha_{s}}{8 \pi v}
  \frac{4m_{q}^{2}}{q_{h}^{2}}
  \left[
  1 + \left( 1 - \frac{4m_{q}^{2}}{q_{h}^{2}} \right)
  f \Bigl( \frac{4m_{q}^{2}}{q_{h}^{2}} \Bigr)
  \right],
\label{Eq:ghgg}
\end{split}
\end{equation}
%%%
where $v = 246\,\text{GeV}$ is the vacuum expectation value 
of the Higgs field, $\alpha_{s} = g_{s}^{2}/4\pi$, and $m_{q}$ 
represents the mass of the running quark. The function $f (r)$ is 
given as, 
%%%
\begin{equation}
\begin{split}
  f \left( r \right)
  = - \frac{1}{4} \log^{2}
  \left[
  - \frac{1 + \sqrt{1-r}}{1 - \sqrt{1-r}} \,
  \right] \ \ \ \ (r<0).
\label{Eq:f}
\end{split}
\end{equation}
%%%
Note that the function $f (x)$ has 
no imaginary part since 
$q_{h}^{2} < 0$ for the $t$-channel scattering $\ell_{i} g \to 
\tau g$. This is different from the on-shell Higgs production at 
the LHC, where the scale is fixed at $q_h^2= m_h^2$, and 
therefore, the lighter quark contributions  induce
the imaginary part.

\begin{figure}[t]
\begin{center}
\includegraphics[width=80mm]{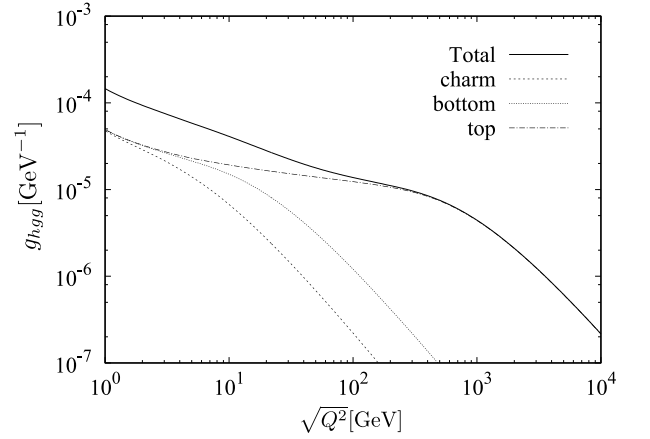}
\vspace{-0.5cm}
\end{center}
\caption{Scale dependence of $g_{hgg}$ and
each quark contribution.}
\label{Fig:ghgg}
\end{figure}

We count only charm, bottom, and top loop contributions for the 
effective coupling. Each loop contribution becomes larger and 
approach to a constant value for smaller $Q^2$ since the loop function 
has the following asymptotic form, 
%%%
\begin{equation}
\begin{split}
 r
  \left[
  1 + \left( 1 - r \right)
  f (r)  \right] \to
 \begin{cases}
  2/3 & (r \to -\infty, Q^2 \to 0).\cr
  0 & (r \to 0 , Q^2 \to \infty).
  \end{cases}
\label{Eq:limit}
\end{split}
\end{equation}
%%%
Therefore, in addition to the top quark loop, 
the relatively heavy quarks ($b, c$)  also 
contribute while the lighter quarks 
($u, d, s$) contributions are still suppressed in the DIS regime 
$Q^2 \gtrsim 1$~(GeV)$^2$. 
%%%
This is different from the case at the LHC, where the dominant 
contribution is essentially only via the top quark loop as $q_h^2=m_h^2$.
Figure~\ref{Fig:ghgg} shows the $Q^2$ dependence of
$g_{hgg}$ and each quark contribution.
%%%
Due to the constructive 
contributions there exists 
a sizable enhancement of the cross sections relative to the case with 
top contribution only. Note that the gluon operator for the CLFV hadronic 
tau decays is derived by integrating out the heaviest three quarks 
since energy scales of those reactions are at $\mathcal{O}(100)\, 
\text{MeV}$~\cite{Celis:2013xja}. 
%%%
The gluon operator derived under the same condition makes the 
correlation of the decays and the scattering $\ell_{i} g \to \tau g$ 
to be clear, and leads to comprehensive understanding of the CLFV 
interactions.

%%%%%%%%%%%%%%%%%%%%%%
%%%%%%%%%%%%%%%%%%%%%%
%%%%%%%%%%%%%%%%%%%%%%
\subsection{cross sections}
\label{Sec:cross}
%%%%%%%%%%%%%%%%%%%%%%
%%%%%%%%%%%%%%%%%%%%%%
%%%%%%%%%%%%%%%%%%%%%%

The Lagrangian~\eqref{Eq:Lag} describes the two types of 
subprocesses, $\ell_{i} g \to \tau g$ (Fig.~\ref{Fig:diagram} (a)) 
and $\ell_{i} g \to \tau q\bar{q}$ (Fig.~\ref{Fig:diagram} (b)). 
The total cross section is formulated as 
%%%
\begin{equation}
\begin{split}
	\sigma_{\ell_{i}N \to \tau X} \!
	= \!\!\! \sum_{\hat{X}=g, q\bar{q}}
	\int \hspace{-1mm} dx dy 
	\int_{0}^{1} \hspace{-1mm} d\xi 
	\frac{d^2\hat{\sigma}_{\ell_{i} g \to \tau \hat{X}}}{dx dy} 
	f_{g}(\xi, Q^{2})\,,
\label{Eq:sigma}
\end{split}
\end{equation}
%%%
where $x$ is the Bjorken variable and 
$y$ is the measure of inelasticity; 
%%%
\begin{equation}
\begin{split}
	x \equiv \frac{Q^{2}}{2P \cdot q_{h}}, ~~ 
	y \equiv \frac{2P \cdot q_{h}}{2P \cdot p_{i}}\,.
\label{Eq:xy}
\end{split}
\end{equation}
%%%
Here, $P$ and $p_{i}$ denote momenta of the initial nucleon 
and the initial lepton. Note that the momentum transfer 
$q_h = p_i - p_{\tau}$ is defined only with the initial lepton 
and the final tau momenta but not with the momentum related 
with $\hat{X}$.
%%%
The ranges of $x$ and $y$ are obtained as~\cite{Albright:1975, 
Hagiwara:2003di}, 
%%%
\begin{equation}
\begin{split}
	&
	x_\text{min} = \frac{m_{\tau}^{2}}{(s-M^{2}) - 2m_{\tau}M}, ~~ 
	x_\text{max} = 1, ~~
	\\[1mm]&
	y_\text{min} = \frac{1}{2}(A-B), ~~ 
	y_\text{max} = \frac{1}{2}(A+B), 
	\\[1mm]&
	A= %\frac{1}{2} 
	\frac{s-M^{2}}{(s-M^{2}) + xM^{2}} 
	\biggl[ 
	1 - \frac{m_{\tau}^{2}}{x(s-M^{2})} 
	- \frac{2 m_{\tau}^{2} M^{2}}{(s-M^{2})^{2}}
	\biggr], 
	\\[1mm]&
	B= %\frac{1}{2} 
	\frac{s-M^{2}}{(s-M^{2}) + xM^{2}}
	\sqrt{
	\biggl( 1 - \frac{m_{\tau}^{2}}{x(s-M^{2})} \biggr)^{2} 
	- \frac{4 m_{\tau}^{2} M^{2}}{(s-M^{2})^{2}}
	}\,.
\label{Eq:min_max}
\end{split}
\end{equation}
%\end{widetext}
%%%
%%%
Here $M$ and $m_{\tau}$ denote the masses of nucleon and 
tau lepton, respectively. The gluon parton 
distribution function (PDF) is denoted as $f_{g}(\xi, Q^{2})$ and $\xi$ is the four-momentum 
fraction of the nucleon carried by the parton, 
$p_{g} = \xi P$. The 
range of $\xi$ depends on the subprocess.

The parton level differential cross section of $\ell_{i} g \to \tau g$ 
in a massless limit of incident lepton is given by
%%%
\begin{equation}
\begin{split}
	&\frac{d\hat{\sigma}_{\ell_{i} g \to \tau g}}{dxdy}
	=
	\frac{Q^{4}\left(Q^{2} + m_{\tau}^{2}\right)}{8 \pi \hat{s}}
	\\& ~~~ \times 
	\frac{|g_{hgg}|^{2}\left( \left| \rho_{i \tau} \right|^{2} 
	+ \left| \rho_{\tau i} \right|^{2} \right)}{(Q^{2} + m_{h}^{2})^{2}} 
	\delta(\xi-x) \,. 
\label{Eq:sigma_lglg}
\end{split}
\end{equation}
%%%
The invariant mass of the system is denoted by $\hat{s} = (p_{i} + 
p_{g})^2$. We have the relation $\xi = x$ as the outgoing gluon is 
massless.
%%%
In the same limit, the parton level differential cross section of 
$\ell_{i} g \to \tau q\bar{q}$ is given by
%%%
\begin{equation}
\begin{split}
	&\frac{d\hat{\sigma}_{\ell_{i} g \to \tau q \bar{q}}}{dxdy}
	=
	\frac{\alpha_{s}y\left(Q^{2} + m_{\tau}^{2}\right)}
		{64 \pi^{2}\xi\left(Q^{2} + w^{2}\right)^{2}}
	\\& ~~ \times  
	\biggl\{ 
	2 K w^{2} (4m_{q}^{2} + Q^{2}) 
	+ \biggl[ 
	(Q^{2} + w^{2})^{2}
	\\& ~~ 
	-2 (4m_{q}^{2} + Q^{2}) (w^{2} - 2m_{q}^{2}) 
	\biggr] 
	\log\biggl| \frac{1+K}{1-K} \biggr|
	\biggr\} \\& ~~
	\times\frac{y_{q}^{2} \left( \left| \rho_{i\tau} \right|^{2} 
	+ \left| \rho_{\tau i} \right|^{2} \right)}
	{\left(Q^{2} + m_{h}^{2}\right)^2} 
	\theta\left(\xi-x\frac{Q^2+4m_q^2}{Q^2}\right),
\label{Eq:sigma_lglqq}
\end{split}
\end{equation}
%%%
where $K \equiv \sqrt{1 - 4m_{q}^{2}/w^{2}}$, and 
$w^{2} = (p_g+q_{h})^{2}=(p_q + p_{\bar{q}})^2 = 
Q^2(\xi/x-1) $ is the invariant mass of the final quark and 
anti-quark system. For $\ell_{i} g \to \tau q \bar{q}$, to 
correct the finite mass effect of the outgoing quarks $m_q$, 
$\xi=x(Q^{2} + w^2)/Q^{2}$ is taken~\cite{Georgi:1976}.

%%%%%%%%%%%%%%%%%%%%%%
%%%%%%%%%%%%%%%%%%%%%%
%%%%%%%%%%%%%%%%%%%%%%
\subsection{general form of CLFV gluon operator}
\label{Sec:eff}
%%%%%%%%%%%%%%%%%%%%%%
%%%%%%%%%%%%%%%%%%%%%%
%%%%%%%%%%%%%%%%%%%%%%

We briefly discuss the CLFV scatterings $\ell_{i} N \to \tau X$ 
mediated by other heavy particles which (in)directly couples 
with the gluon field, for example, heavy Higgses $H$ and $A$ in two Higgs 
doublet models.
As long as those particles are heavy enough we can describe 
the subprocess $\ell_{i} g \to \tau g$ using the following 
effective operators,
%%%
\begin{equation}
\begin{split}
	\mathcal{L}_\text{eff} 
	= \overline{\ell_j}\left(C_{ij}P_L+C_{ji}P_R\right) \ell_i 
	G_{\mu\nu}^aG^{a\mu\nu}.%+h.c.
\label{Eq:contact}
\end{split}
\end{equation}
%%%
The constraints on $C_{ij}$ come from the searches for 
CLFV tau decays, $\text{BR}(\tau \to \mu \pi^+\pi^-) < 2.1 
\times 10^{-8}$ and $\text{BR}(\tau \to e \pi^+\pi^-) < 2.3 
\times 10^{-8}$~\cite{Miyazaki:2013}, which are corresponding 
to $|C_{\mu\tau}|^2 + |C_{\tau\mu}|^2 < 7.40 \times 
10^{-18}$~GeV$^{-6}$ and $|C_{e\tau}|^2 + |C_{\tau e}|^2 
< 8.10\times 10^{-18}$~GeV$^{-6}$~\cite{Celis:2014asa}.
%%%
Note that these constraints are much weaker than 
those assuming only via the SM Higgs,
where the stringent constraints on $\rho_{ij}$ is applicable through
the relation $C_{ij} = \rho_{ij} g_{hgg} /m_h^2$.

The differential cross section of the subprocess $\ell_{i} g \to \tau 
g$ is easily calculated by ignoring $Q^2$ term in the denominator 
in Eq.~\eqref{Eq:sigma_lglg}, 
%%%
\begin{equation}
\begin{split}
	&\frac{d\hat{\sigma}_{\ell_{i} g \to \ell_{j} g}^\text{contact}}{dxdy}
	= \frac{Q^4\left(Q^2+m_j^2\right)}{8\pi\hat{s}}
	\left(\left|C_{ij}\right|^2+\left|C_{ji}\right|^2\right)\delta(\xi-x).
\label{Eq:sigma_contact}
\end{split}
\end{equation}

%%%%%%%%%%%%%%%%%%%%%%
%%%%%%%%%%%%%%%%%%%%%%
%%%%%%%%%%%%%%%%%%%%%%
\section{Numerical analysis}
\label{Sec:num}
%%%%%%%%%%%%%%%%%%%%%%
%%%%%%%%%%%%%%%%%%%%%%
%%%%%%%%%%%%%%%%%%%%%%

\begin{figure}[t]
\begin{center}
\includegraphics[width=80mm]{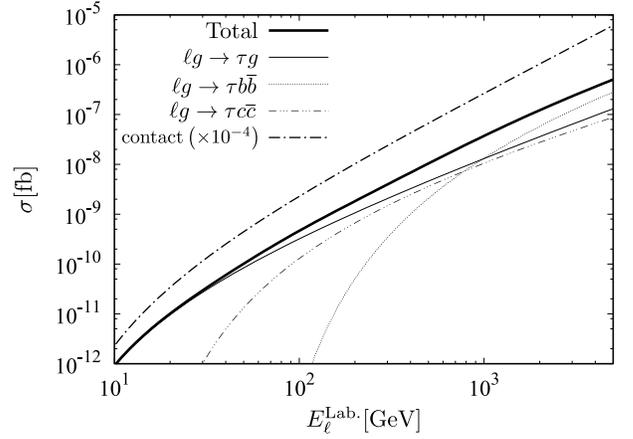}
\end{center}
\vspace{-0.5cm}
\caption{Cross sections of $\ell_{i} N \to \tau X$ as 
a function of incident lepton beam energy for 
$\sqrt{|\rho_{\ell \tau}|^2+ 
|\rho_{\tau \ell}|^2}= 2.4 \times 10^{-3}$. The dot-dashed line shows the cross 
section for the effective interaction~\eqref{Eq:contact} for
$|C_{\ell\tau}|^2 + |C_{\tau \ell}|^2 
= 8.10\times 10^{-18}$~GeV$^{-6}$
and other lines show total and partial cross sections of the 
scatterings mediated by the SM Higgs. }
\label{Fig:low}
\end{figure}

\begin{figure}[t]
\begin{center}
\includegraphics[width=80mm]{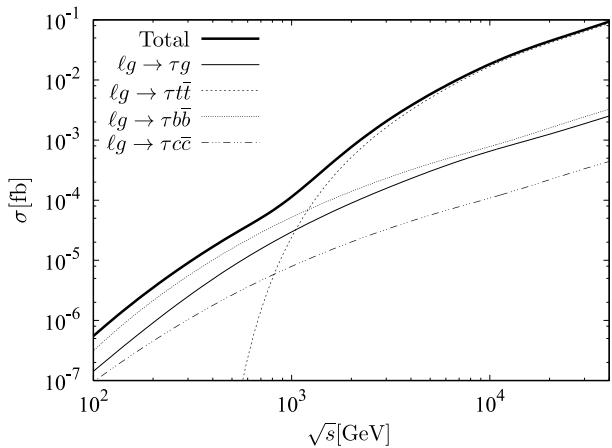}
\end{center}
\vspace{-0.5cm}
\caption{Cross sections of Higgs mediated scattering 
$\ell_{i} N \to \tau X$ for beam collision experiments. 
We take maximally allowed CLFV Yukawa couplings. 
Thick solid line shows total cross section, and other 
lines are partial cross sections. }
\label{Fig:high}
\end{figure}

We perform a numerical analysis. 
In our analysis, the maximally allowed CLFV Yukawa coupling for electron,
$\sqrt{|\rho_{e \tau}|^{2} + |\rho_{\tau e}|^{2}} 
= 2.4 \times 10^{-3}$~\cite{Khachatryan:2016rke} is taken.
%%%
Other input parameters are  
$m_{h} = 125.7\,\text{GeV}$, 
$m_{\tau} = 1.777\,\text{GeV}$, 
$m_{t} = 173.2\,\text{GeV}$, 
$m_{b} = 4.180\,\text{GeV}$, 
$m_{c} = 1.275\,\text{GeV}$, and 
$M = 0.9383\,\text{GeV}$~\cite{Olive:2016xmw}. 
%%%
We have used CTEQ6L1 for the nucleon PDF~\cite{Pumplin:2002}. 
We restrict the phase space integration with $W^2>(1.5~{\rm GeV})^2$ 
and $Q^2>(1~{\rm GeV})^2$ to ensure that the parton model picture is valid, 
where $W^{2} = (P+q_{h})^{2}$ is the hadronic invariant 
mass. We consider only the DIS regime and ignore other resonant 
effects which are known to be sub-dominant~\cite{Hagiwara:2003di}.

%%%%%%%%%%%%%%%%%%%%%
%%%%%%%%%%%%%%%%%%%%%
%%%%%%%%%%%%%%%%%%%%%
\subsubsection{fixed target experiments}
%%%%%%%%%%%%%%%%%%%%%
%%%%%%%%%%%%%%%%%%%%%
%%%%%%%%%%%%%%%%%%%%%

First we consider the prospects at fixed target experiments.
Figure~\ref{Fig:low} shows the cross section of the Higgs 
mediated scattering $\ell_{i} N \to \tau X$ as a function of 
the incident lepton beam energy.  
%%%
We show the contributions from each subprocess 
separately, and also the sum in a thick solid line. 
Due to the large gluon PDF and no phase space suppression, 
the new subprocess $\ell_{i} g \to \tau g$ 
leads to the large enhancement in $\sigma (\ell_{i} N \to \tau X)$.
%%%
The ratio between $\sigma (\ell_{i} N \to \tau X)$ with and 
without the new subprocess is 7.8 (1.8) for $E_{\ell} = 50\, 
\text{GeV}$ ($500\,\text{GeV}$). The subprocess 
$\ell_{i} g \to \tau c\bar{c}$ ($\ell_{i} g \to \tau b\bar{b}$) 
only starts to be relevant at $E_\ell \sim 100$\,GeV 
($500$\,GeV) due to the phase space suppression by the 
production of a pair of quark and anti-quark. 
%%%
Therefore, for $E_\ell \lesssim 1$\,TeV, the 
dominant contribution for the CLFV scattering in fixed target 
experiments comes from the new subprocess. 
%%%
Inclusion of the $g_{hgg}$ coupling enhancement shown in 
Fig.~\ref{Fig:ghgg} is also important. Typically, it provides a 
factor of $3\sim 7$ enhancement relative to the case with
top contribution only.

Event rate of $\ell_{i} N \to \tau X$ process at the fixed target 
experiments is estimated as~\cite{Abada:2016vzu}, 
%%%
\begin{equation}
\begin{split}
	N 
	\simeq 
	6 \times 10^{-16} \cdot N_{\ell_{i}} 
	\left( \frac{\sigma_{\ell_{i}N \to \tau X}}{1\,\text{fb}} \right) 
	\left( \frac{T_{m}}{1\,\text{g}\,\text{cm}^{-2}} \right),
\label{Eq:events}
\end{split}
\end{equation}
%%%
where $N_{\ell_{i}}$ is the intensity of $\ell_{i}$ per year, and 
$T_{m}$ is the target mass in unit of $\text{g}\,\text{cm}^{-2}$. 
According to the formula, $\mathcal{O} (10)$ ($\mathcal{O} 
(10^{3})$) events of $\ell_{i} N \to \tau X$ are expected per year 
for the electron beam energy of an upgrade option in ILC (PWFA), 
$E_{e} = 500\,\text{GeV}$ ($5\,\text{TeV}$),
$T_{m} = 100\,\,\text{g}\,\text{cm}^{-2}$, and the electron 
intensity $N_e= 10^{22}/$year. 

The scattering cross section assuming the contact operator 
with the maximally allowed value is shown in a dot-dashed line.
For muon options, 
we require the neutrino factories,
which would reach at $N_\mu=10^{20}$/year~\cite{Delahaye:2013jla}
with a beam energy of $\mathcal{O}(100)$~GeV to 
provide $\mathcal{O} (10)$ events/year.
The currently available intensity is not enough, for example,
$N_\mu=10^{15}$/year in COMPASS II 
experiment~\cite{Gautheron:2010wva}
and $N_\mu=10^{19}$/year with a lower beam energy in 
COMET~\cite{Cui:2009zz} in future. 

%%%%%%%%%%%%%%%%%%%%%
%%%%%%%%%%%%%%%%%%%%%
%%%%%%%%%%%%%%%%%%%%%
\subsubsection{collider experiments}
%%%%%%%%%%%%%%%%%%%%%
%%%%%%%%%%%%%%%%%%%%%
%%%%%%%%%%%%%%%%%%%%%

Next, we turn to the prospects at collider experiments. 
The total cross section of the Higgs mediated scattering as a function of the collision energy 
is shown in Fig.~\ref{Fig:high}. 
It is essentially the same quantity shown in Fig.~\ref{Fig:low} but in different regime
since the collision energy at fixed target experiments is related with 
$\sqrt{s}\simeq \sqrt{2ME_{\ell}} \sim 1.4~{\rm GeV} 
\sqrt{E_{\ell}/{\rm GeV}}$.
%%%
As the collision energy $\sqrt{s}$ increases the cross section grows rapidly.
When it reaches at $\sim 2m_{t}$, a subprocess 
$\ell_{i} g \to \tau t\bar{t}$ starts to contribute and becomes dominant for 
$\sqrt{s} \gtrsim 1\,\text{TeV}$.
%%%

A future electron-proton beam collision experiment TLHeC (VHE-TLHeC) 
plans to achieve a $\sqrt{s} \simeq 1.3\, (3.5) \, \text{TeV}$ with 
a luminosity of $\sim 10^{35}\,\text{cm}^{-2} \text{s}^{-1}$, i.e., 
$\mathcal{O}(1000)\,\text{fb}^{-1}/$year~\cite{Zimmermann}. 
We expect $\mathcal{O}(100)$ events for the maximal allowed CLFV coupling.
%%%

Although such a high energy provides an opportunity to discover 
$\ell_{i} N \to \tau X$, it is challenging to identify 
the SM Higgs as the CLFV mediator due to the 
high dominance of $\ell_{i} g \to \tau t\bar{t}$.
By measuring each cross section of the subprocess
both at the fixed target and at the beam collision experiments, 
we can comprehensively understand the CLFV interactions 
with the search results for CLFV tau decays including the 
gluon operator. It could also judge whether the CLFV 
scattering is mediated by the SM Higgs or the other particles. 

Experimental feasibility of measuring each subprocess should be also
studied further and we can study
e.g., the angular distributions of the tau and the hadronic
system $X$, the compositions of the final states, and so on. 
%%%
We leave these issues for our future works.

%%%%%%%%%%%%%%%%%%%%%
%%%%%%%%%%%%%%%%%%%%%
%%%%%%%%%%%%%%%%%%%%%
\section{Summary and Outlook} 
\label{Sec:sum}
%%%%%%%%%%%%%%%%%%%%%
%%%%%%%%%%%%%%%%%%%%%
%%%%%%%%%%%%%%%%%%%%%

In summary, the Higgs mediated CLFV scattering 
$\ell_{i} N \to \tau X$ has been reconsidered 
by taking 
(i) new subprocess $\ell_{i} g \to \tau g$, and 
(i\hspace{-0.5pt}i) sea quark number conserving 
subprocess $\ell_{i} g \to \tau q\bar{q}$ ($q = c, b, t$), 
into account. 
%%%
At the fixed target experiments with $E_{\ell} \lesssim 1\,
\text{TeV}$, the total cross section $\sigma(\ell_{i} N \to 
\tau X)$ is enhanced more than about twice by the 
subprocess $\ell_{i} g \to \tau g$. 
%%%
It has been pointed out that the associated quarks are only produced 
in pairs in $\ell_{i} N \to \tau X$, 
and hence $\sigma(\ell_{i} N \to \tau q \bar{q})$ 
starts to be relevant on higher beam energy than that estimated 
in the previous works wherein the 
phase space suppression by a quark pair production is not considered. 
%%%
$\mathcal{O}(10) - \mathcal{O}(10^{3})$ events of 
$\ell_{i} N \to \tau X$ could be expected in both the fixed 
target experiments and the beam collision experiments in future, 
and it would provide opportunities to shed light on the nature of 
Higgs mediated CLFV complementarily with other experiments, 
such as CLFV tau decay searches. 
%%%
Our results hold for other CLFV mediators which
couple with gluon and/or heavy quarks.

%%%%%%%%%%%%%%%
%%%%%%%%%%%%%%%
\section*{Acknowledgments}
%%%%%%%%%%%%%%%
%%%%%%%%%%%%%%%

We thank Y. Kiyo, S. Kumano, M. Nojiri, K. Tobe, and H. Yokoya
for useful discussions and valuable comments.
This work is supported in part by the JSPS Grant-in-Aid for
Scientific Research Numbers JP16H03991, JP16H02176, 17H05399 (M.T.), 
16K05325 and 16K17693 (M.Y.).
MT is supported by World Premier International Research 
Center Initiative (WPI Initiative), MEXT, Japan.

%%%%%%%%%%%%%%%
%%%%%%%%%%%%%%%

\end{document}